\documentclass[12pt]{article}
\usepackage{graphicx}
\usepackage{cite}
\usepackage{times}
\usepackage{xspace}
\usepackage{geometry}
\usepackage[utf8]{inputenc}
\usepackage{enumitem,amssymb,wrapfig,natbib}
\usepackage{ragged2e}
\newlist{thematic}{itemize}{8}
\setlist[thematic]{label=$\square$}
\usepackage{pifont}

\newcommand{\SII}{\ion{S}{II}\xspace}
\newcommand{\Halpha}{${\rm H}\alpha$\xspace}
\newcommand\ion[2]{#1$\;${\scshape{#2}}}

\begin{document}
\raggedright
\huge
Astro2020 Science White Paper \linebreak

Making the Connection between Feedback and Spatially Resolved Emission Line Diagnostics \linebreak
\normalsize

\noindent \textbf{Thematic Areas:} $\square$ Star and Planet Formation \hspace*{20pt}\linebreak
$\square$ Resolved Stellar Populations and their Environments \hspace*{40pt} \linebreak
$\square$    Galaxy Evolution\linebreak
  
\textbf{Principal Author:}

Name: Eric W.\ Pellegrini
 \linebreak						
Institution:  University of Heidelberg
 \linebreak
Email: eric.pellegrini@uni-heidelberg.de
 \linebreak
Phone:  +49 6221 54 6713
 \linebreak
 
\textbf{Co-authors:} N.\ Drory (McDonald Observatory, UT Austin), Guillermo A.\ Blanc (Carnegie), Juna A.\ Kollmeier (Carnegie), Sarah E.\ Tuttle (University of Washington), Laura A.\ Lopez (The Ohio State University), Josh Simon (Carnegie), Amy M.\ Jones (The University of Alabama), Vladimir Avila-Reese (UNAM), Kathryn Kreckel (Max Planck Institute for Astronomy), Renbin Yan (University of Kentucky) 
\vspace*{3mm}

\textbf{Abstract:}
\justifying
Crucial progress in our understanding of star formation and feedback will depend on the ability to obtain spatially resolved spectroscopic observations of \ion{H}{ii} regions, from which reliable instantaneous measurements of their physical conditions can be obtained. Acquiring these datasets across full galactic systems will prove crucial for obtaining population samples that enable us to understand the time evolution of similar regions, and the variability of conditions among coeval regions. Separating the spatial and temporal dependencies in such way for different physical processes involved in star formation and the injection of feedback is crucial to overcome the inherit degeneracies associated with observing instantaneous snapshots of a dynamic ISM at any given time. Emission line diagnostics are at the core of measuring the physical condition in \ion{H}{ii}  regions (e.g. dynamics, SFR, chemical abundances, dust extinction, ionization and excitation, etc.). These measurements require high spatial resolution, contiguous coverage across full galactic systems, and sensitivities significantly deeper than past efforts. The spatial scale required to resolve the \ion{H}{ii} regions of a few pc is only attainable in the Local Group where very large sky coverage is necessary.

\pagebreak

\section{Introduction}

Nebulae associated with star formation are prodigious emitters of line emission in galaxies. On scales of $\approx 10 - 50$~pc, relatively dense \ion{H}{ii} regions reprocess much of the emitted ionizing stellar radiation, as evidenced by their large contribution to the integrated galactic \Halpha luminosity. These dense nebulae provide the primary barrier to radiative heating of the diffuse ISM, as well as the escape of stellar winds and shocked gas. There is however ample evidence that this barrier is often breached \citep{Haffner2009, Seon2009} in systematic ways depending on variables like environment or cluster luminosity \citep{beckman2000AJ}.

\hspace*{10mm}It is now clear that star formation must be a self-regulated process: gas collapses to form stars, which provide energetic feedback in the form of radiation and stellar winds to the surrounding gas, heating it and thereby regulating the subsequent birth of more stars. Despite its importance, feedback is not understood \textbf{\textit{quantitatively}}: how strongly, how far-reaching, and for how long does it affect its environment, and how this changes with stars of different masses, ages and metal abundances? 

\hspace*{10mm}We argue in this White Paper that crucial progress in the coming decade will depend on the ability to obtain spatially resolved spectroscopic observations of \ion{H}{ii} regions for reliable instantaneous measurements of their physical conditions, with full galactic coverage to obtain a population sample that enables us to understand the time evolution of similar regions as well as the variability of conditions among coeval region. Further, these measurements require sensitivities significantly deeper than past efforts.

\section{What is the challenge in ``observing'' feedback?}
\label{sect:method}

\begin{wrapfigure}{R}{9cm}
    \vspace*{-0.5cm}
    \includegraphics[scale=0.4]{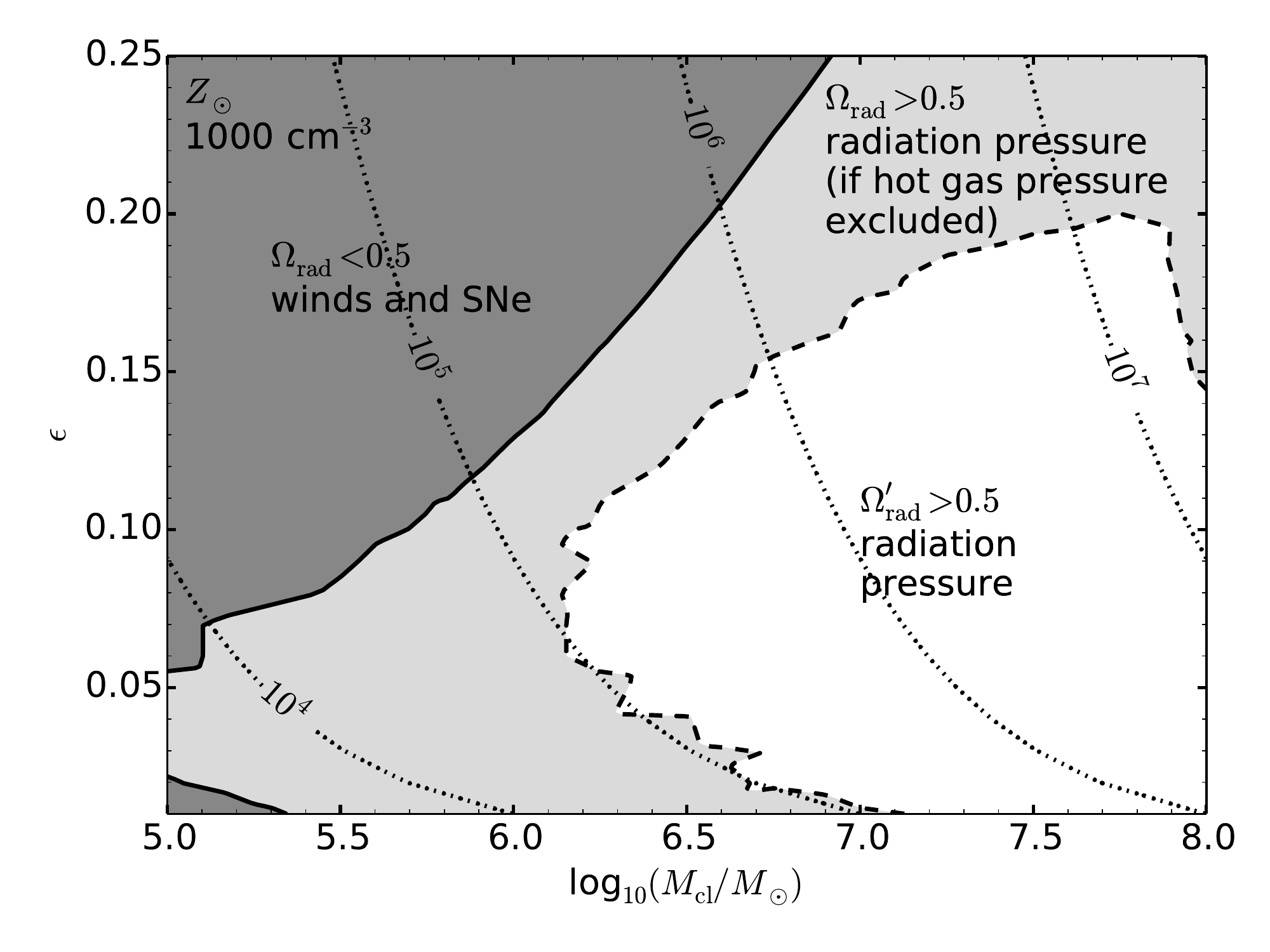}
    \caption{\footnotesize As a consequence of including all significant stellar feedback mechanisms and the effects of gravity on clouds of finite mass, the resulting dynamics have no closed solution. To first order, the dominance of a feedback mechanism in the net kinetic energy of the system depends on star formation efficiency (SFE) and cluster mass.}
    \label{fig:ForceRegime}
\end{wrapfigure}

Unfortunately it is not possible to observe the evolution of an individual star forming region, we can only measure the current emission spectrum and extrapolate the history of its dynamical evolution. Even in geometrically simple models which include all feedback forces, gravity, and their relative coupling, evolution is largely degenerate. A non-linearity of feedback interactions and cooling result in different feedback forces dominating under different circumstances at different times. For example, winds may dominate forces at early times by creating hot bubbles from thermalized shocks. The bubbles then cool, resulting in x-ray faint objects with little present day driving by hot gas, even if the acceleration by such hot gas dominates the total net kinetic energy of the system. Stellar radiation pressure from ionizing radiation is large at early times when gas densities are high, but then expansion lowers the gas densities until a significant fraction of the gas is fully ionized, and the radiation becomes decoupled. Some issues have become simpler, with infrared pressure falling out of favor as a significant force when spectral shifting is properly treated \citep{Reissl2018}. The remaining parameter space that determines the evolution of each object is cluster mass, environmental density and metallicity.  Together these dictate which feedback mechanism dominates, such as in Figure \ref{fig:ForceRegime}. {\bf Thus a large uncertainty for the next decade is not what is the physics of feedback, but how does nature sample the myriad of combinations of cloud and cluster conditions within and between galaxies.}

The challenges in making quantitative estimates of feedback stem from the need to constrain two physical properties only indirectly constrained by observations: The engines which drive evolution (the stellar content, its age, and all properties related to feedback), and the structure of the ISM which is not only driven, but determines how different feedback processes couple to the ISM.


\hspace*{10mm}An important advancement in the connection of observation to theory came in \cite{BFM91}. Using optical spectroscopy to constrain the line of sight distance of the Orion nebula along a long-slit observation, it was found that the gas density at the ionization front, as measured from the \SII doublet, correlated strongly with the 3d distance of the nebula from the star cluster. Later studies would reveal that such a correlation is the result of a quasi-hydrostatic equilibrium between feedback forces of winds and radiation and their interaction with the ISM through shocks and radiative transfer. These results showed that stellar feedback and the ISM were not only related, but feedback rapidly sets the instantaneous conditions across the $H^+/H_0/H_2/CO$ interfaces, mapping present day emission spectra to present day feedback. This is both a blessing, and a curse. It means that we can make a direct connection between current stellar populations and their feedback to ISM conditions, but must rely on theory and ensembles of objects drawn from well sampled populations to understand the past cumulative effects of a feedback-and-evolution history, and predict its future.

\section{ISM Structure and Emission}
Decades of research has shown that ISM emission line diagnostics are intricately related to the structure of the ISM around massive stars. These provide useful information on quantities indirectly related to feedback, key among them the dimensionless ionization parameter $U$, defined as 
\begin{equation}
	\label{eq:U}
    U = \frac{Q_{0}}{4\pi r^2 n(H) c}
\end{equation}
where $Q_0$ is H ionization rate photons produced in a stellar population,  $r$ is the geometric distance of a piece of the ISM, with density $n(H)$, from an ionization source. For \ion{H}{ii} regions which expand with even modest winds from the lowest mass O-stars, the thickness of the photo-ionized region of the shell $\Delta r$ is relatively thin compared to $r$. A move beyond qualitative description of feedback with assumptions about gas distribution \citep{2011ApJ...731...91L, Lopez2014} is possible by connecting $U$ and stellar feedback in {\em spatially resolved} nebulae where radiation pressure locally can be ideally reformulated as 
\begin{equation}
    P_{rad} = U \times (h\bar{\nu}_{ion}~n(H))
\end{equation}

\hspace*{10mm}It is important to note that interior to the \ion{H}{ii} region is the shocked wind bubble with temperatures $T\gtrsim 10^6$~K. Being of low density and collisionally ionized, it does not absorb stellar radiation, but provides a source of thermal pressure. The surface brightness of X-ray observations of diffuse gas provide direct measures of bubble pressure via
\begin{equation}
	F_{X,bol} = \Lambda n_{e,X}^2 \eta \times dl_{los}
\end{equation}
where the line of path length $dl_{los}$ is related to the observed X-ray flux $F_{X,bol}$ by the electron density $n_{e,X}$ and the cooling function $\Lambda$ \citep{Townsley2003}, and is insensitive to whether that pressure originates from winds and/or supernovae.

\hspace*{10mm}Here we have made the case that quantitative measurements of feedback by winds and radiation hinge on creating a bridge between stellar sources and the geometry of a region via its spatially resolved observable emission. To make meaningful maps of pressure variations, spatial resolutions that can separate the \ion{H}{II} from a potential X-ray bubble, or a few parsecs, are needed. These scales are presently only reachable in the Local Group, however implying very large sky coverage.

\hspace*{10mm}This approach requires solving for unknowns such as metal and stellar content from emission lines, as demonstrated in \cite{Pellegrini2011}. Unlike in unresolved objects, spatially resolved nebula open new doors for studying feedback. The relative spatial-distribution of high- and low-ionization emission, such as [\ion{S}{II}] and [\ion{O}{II}] depend primarily on three fundamental properties characterizing \ion{H}{ii} regions: (1) Stellar ionizing SED, luminosity, (2) metallicity, (3) the 3 dimensional distribution of gas in velocity and density. In star forming regions dominated by relatively simple stellar populations, the first two parameters can safely be assumed to be constant across the nebula. Thus variations in the ionization parameter, $U$, are driven by changes in the volume density and geometric distance of gas to stars. The variations in line ratios with $U$ trace distinct ``paths'' in diagnostic space, dependent on these global parameters. Spatial resolution allows the individual ``path'' to be identified, breaking degeneracy between global parameters and local conditions, making changes in $U$ a powerful probe of 3d structure \citep{Pellegrini2011}. 

\hspace*{10mm}This path will also see assumptions about \ion{H}{ii} regions as idealized, homogeneous objects replaced with variations along individual lines of sight that, which taken together, represent a highly variable and structured ISM.

%

\subsection{Constraining Stellar Sources}
With the advent of missions like Gaia, and the SDSS, mapping of stellar sources across the Milky Way will provide accurate spectral types and effective temperatures millions of stars. However, this is not enough to constrain photoionization models, which are sensitive to parts of the SED impossible to probe on Earth. Emission line spectra provide strong constraints on spectral SEDs on account of the many different ionization states of the ISM they probe \citep[e.g.][]{VilchezPagel1988, Stoy1933, DorsCopetti2003, Zastrow2013}, with dependencies on secondary parameters. The \ion{He}{I}~5876/H$\beta$ ratio provides tight constrains over a large range of stellar effective temperatures and stellar atmosphere, with little dependence on uncertainties in optical depth or gas phase metallicity \citep{Pellegrini2011, Zastrow2013}. With additional constraints provided directly by spectra of stars or clusters, future surveys will provide tight constraints on ionizing sources only available to isolated studies of the past. 

\subsection{Constraining Metal abundances}

The relative strengths of emission lines is highly sensitive to metal abundance. Some species provide dominate cooling channels and are sensitive to total heating rates, while others are sub-dominate and vary with metal abundance. Which channels dominate depend on the stellar effective temperature, which we take to be constrained above. However, to ensure the ability to constrain a wide range of objects with different ages and masses, it is necessary that future surveys cover multiple numerous ionic species in multiple transition states, to ensure an unbiased survey of feedback measurements. Systematics in such methods will be understood independent of photoionization modeling by obtaining abundances directly where auroral line measurements are possible. 


\begin{figure*}
    \centering
    \includegraphics[scale=0.75]{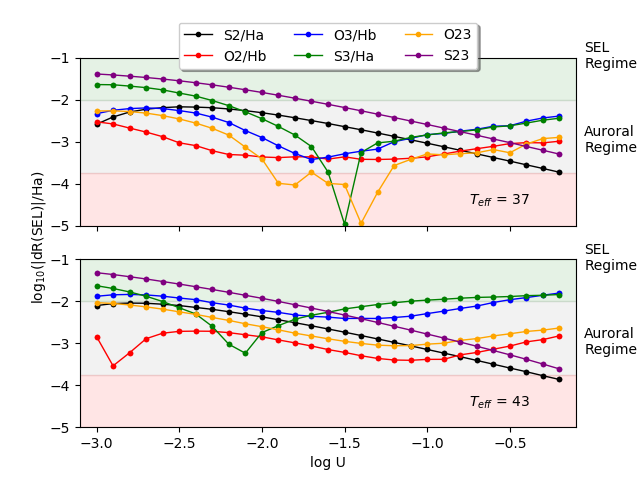}
    \caption{\footnotesize Required line sensitivity per spatial element to resolve changes in ${\rm log_{10}}U = 0.1$, corresponding to an accuracy in 3d-radius of $\approx 10\%$. Less sensitivity maybe required at specific values of U, or specific $T_{eff}$. However, as unbiased surveys do not target individual regions, the full range of the parameters space nature probes must be considered.}
    \label{fig:strong_lin_sensitivies}
\end{figure*}

\section{Revising Strong-line Sensitivities to Constrain Feedback}

The overall approach to breaking the multi-dimensional degeneracy between density, geometry, abundance and stellar SED with spectra is a viable approach to studying feedback. To be applied to large surveys a revision to line sensitivity is required. Strong emission lines typically have intensities relative to ${\rm H}\alpha$ (or ${\rm H}\beta$) of 1 to 2 dex lower, including high ionization [\ion{O}{III}] and [\ion{S}{III}], and low ionization [\ion{O}{II}], [\ion{N}{II}], [\ion{S}{II}]. So called faint emission lines are yet another 1- to 2-dex fainter and require significantly deeper integration times. 

\hspace*{10mm}Typically, observations targeting \ion{H}{ii} regions make a compromise between mapping strong lines at high spatial resolution, and detecting faint lines with spectral averaging across the nebulae for direct measurements of gas phase metal abundances. In the case of the strong lines, the observational objective is phrased as a requirement to observe some fiducial line flux or ratio with the goal of comparison to other objects, or models. To deconstruct spatially resolved physical parameters what is needed is sufficient sensitivity to distinguish expected internal variations in line ratios.

\hspace*{10mm}Take for example, an attempt to distinguish fractional changes in radius of $\Delta r/r \approx 10\%$ across a nebula, assuming we have constrained the other global parameters. From the derivative of Eq.~\ref{eq:U}, one can show that 
\begin{equation}
\label{eq:dr_dU}
|\Delta r/r| = 1/2~ {\rm log_{10}} \Delta U 
\end{equation}
In Fig.~\ref{eq:dr_dU} we show a grid of photoionization models using CLOUDY \citep{Ferland2013} for two different stellar effective temperatures at solar metallicity. Here show a set of emission line ratios most sensitive to structure at intervals of $0.1~{\rm log_{10}} U$, corresponding to the desired accuracy in $\Delta r/r \approx 10\%$. Using the gradient between these models we calculate the needed 3$\sigma$ sensitivity relative to ${\rm H}\alpha$ needed to distinguish one value of $U$ from $U+dU$. \\ 
Our conclusion is that an unbiased survey of \ion{H}{ii} region structure must include a wide range of optical emission lines, spanning from $370$~nm to $1$~$\mu$m, and have at a minimum surface brightness sensitivity in the auroral regime (below $10^{-2} \times {\rm H}\alpha$).

\hspace*{10mm}Nearby galaxies provide populations to sample with sufficient resolution. Investing in infrastructure to allow IFU spectroscopy of large portions of the sky in the Local Group is crucial for obtaining spatially resolved observations of full populations of star forming regions at resolutions $<10$~pc. At the same time, ELTs will enable the study of further away systems presently observable at 50-100~pc with existing facilities, to enlarge the range of physical conditions of star formations we can directly resolve and observe.


%

\begin{wrapfigure}{R}{9cm}
	\vspace*{-0.5cm}
	\includegraphics[scale=0.4]{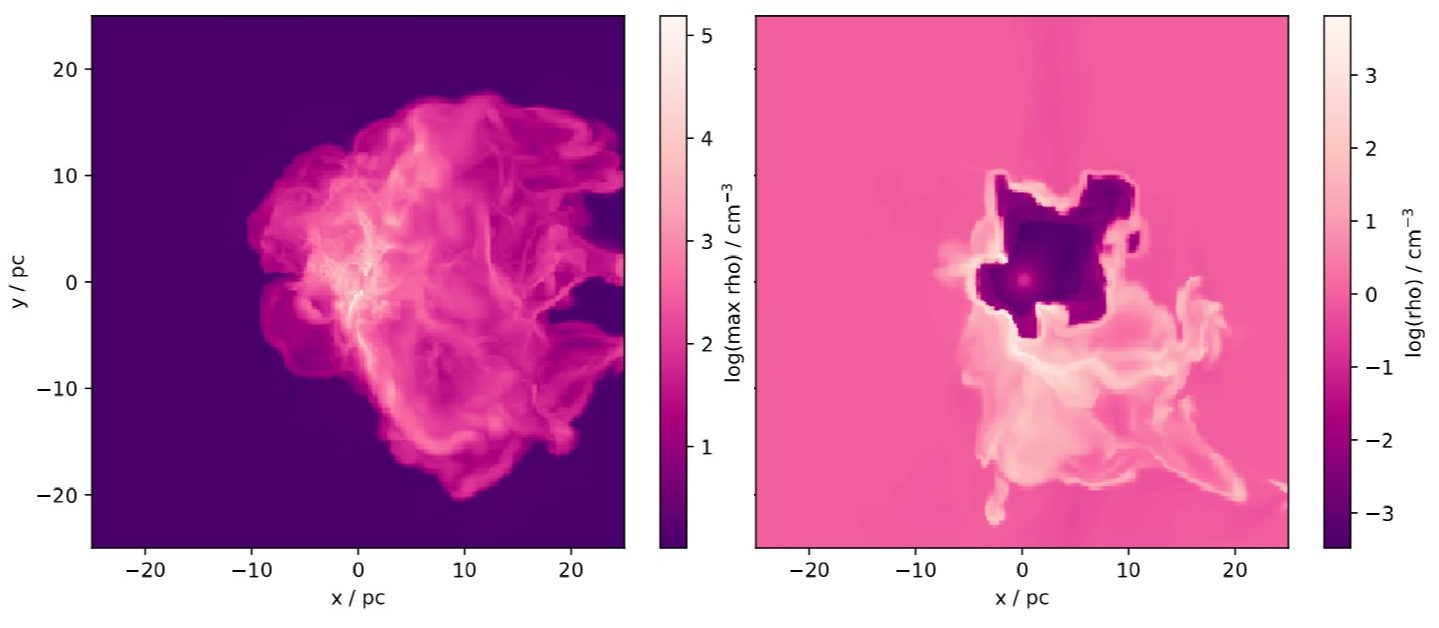}
	\caption{\footnotesize 3d RMHD simulation including stellar winds with the code \textsc{Ramses}. These have sufficient resolution to resolve the feedback structures inside molecular clouds. See \protect\cite{Geen2015b} and \protect\cite{Geen2018} for more details.}
	\label{fig:3d}
\end{wrapfigure}

Cutting edge radiative transfer codes incorporating all the dominate feedback mechanisms: winds, radiation as ionization and momentum, provide the necessary sanity checks on methods seeking to convert emission line diagnostics to quantitative measures of feedback. To test if the underlying assumptions outlined in Sec. \ref{sect:method} the future will turn to 3d simulations with feedback. Radiation (magneto-)hydrodynamic (RMHD) simulations are increasingly sophisticated \citep[][among others]{Dale2005,Gritschneder2009,Peters2010,Walch2012,Colin2013,Howard2016,Geen2018,Ali2018}, but also increasingly expensive. While the resulting morphologies are more complex (Fig.\ref{fig:3d}-left) than analytic approaches like WARPFIELD \citep{Rahner2017}, which are quasi-3d, these 3d simulations still produce lines of sight with emission dominated by 2 density components (Fig.\ref{fig:3d}-right), making the photoionization modeling approach to feedback outlined here a promising method. 

\pagebreak
\begingroup
    \bibliographystyle{apj}
    \setlength{\bibsep}{0pt}
    \bibliography{localbib.bib}
\endgroup

\end{document}